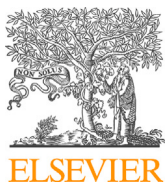
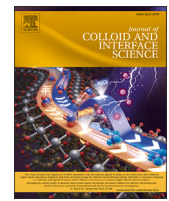

Regular Article

# Electric field-induced clustering in nanocomposite films of highly polarizable inclusions

Elshad Allahyarov [a,b,c,*], Hartmut Löwen [a]

[a] Institut für Theoretische Physik II: Weiche Materie, Heinrich-Heine Universität Düsseldorf, Universitätstrasse 1, 40225 Düsseldorf, Germany
[b] Theoretical Department, Joint Institute for High Temperatures, Russian Academy of Sciences (IVTAN), 13/19 Izhorskaya street, Moscow 125412, Russia
[c] Department of Physics, Case Western Reserve University, Cleveland, OH 44106-7202, United States

GRAPHICAL ABSTRACT

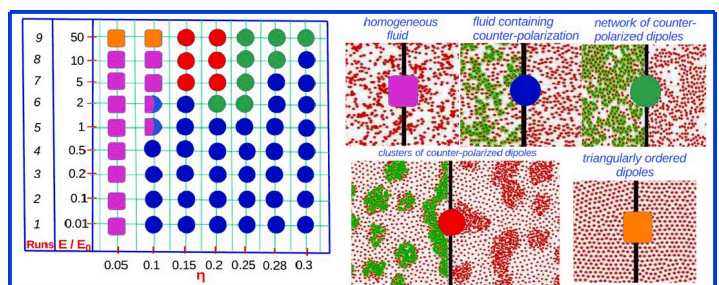

Through computer simulations employing a linearized polarization model, a diverse array of equilibrium clustering of induced dipoles within 2D nanocomposite films is identified. Specifically, stripe and bubble like clusters were detected at high values of the applied field and particle density.

ARTICLE INFO

ABSTRACT

A nanocomposite film containing highly polarizable inclusions in a fluid background is explored when an external electric field is applied perpendicular to the planar film. For small electric fields, the induced dipole moments of the inclusions are all polarized in field direction, resulting in a mutual repulsion between the inclusions. Here we show that this becomes qualitatively different for high fields: the total system self-organizes into a state which contains both polarizations, parallel and antiparallel to the external field such that a fraction of the inclusions is counter-polarized to the electric field direction. We attribute this unexpected counter-polarization to the presence of neighboring dipoles which are highly polarized and locally revert the direction of the total electric field. Since dipoles with opposite moments are attractive, the system shows a wealth of novel equilibrium structures for varied inclusion density and electric field strength. These include fluids and solids with homogeneous polarizations as well as equilibrium clusters and demixed states with two different polarization signatures. Based on computer simulations of an linearized polarization model, our results can guide the control of nanocomposites for various applications, including sensing external fields, directing light within plasmonic materials, and controlling the functionality of biological membranes.

* Corresponding author.
 *E-mail address:* elshad.allahyarov@case.edu (E. Allahyarov).

## 1. Introduction

The need to build adaptive and low-energy expenditure intelligent systems that can be controlled by external stimuli has been promoted by recent advances in sensor technology [1–6]. Among various nanoscale sensing techniques [7], field-effect transistors with channels made of two-dimensional (2D) materials are gaining attention due to their fast response, easy operation, and capability of integration. 2D nanomaterials are also being considered for fabricating wearable, flexible devices of small size for energy scavenging [8]. To improve the capacitive and energy scavenging properties of 2D nanomaterials, they are usually blended with high-$k$ inclusions. In order to fight ionic loss currents in such 2D nanocomposites [9], there is a tendency to make them miniature and a one-layered film.

The basic field-sensing ability of 2D nanomaterials can be understood in two ways: either as the change in its surface area, or as a change in the distribution of the inclusions under poling condition. In the former case, if the poling field is perpendicular to the film surface and the matrix is an elastomer, the induced dipole-dipole repulsion causes a lateral expansion of the film area. In the latter case, strong correlations between induced dipoles might cause an uneven distribution within the matrix, leading eventually to a segregation of the system into regions of low density (gas) and high density (liquid). This phenomenon is the central focus of current work.

Understanding the reorganization of inclusion into clusters and controlling their movement on 2D films are also crucial for comprehending the biomechanical changes in cell membranes. Phospholipids are recognized for possessing a dipole potential, arising from charge accumulation on membrane surfaces and the alignment of dipolar residues and membrane components [10–13]. These strong dipolar fields, ranging from 100 MV/m to 1000 MV/m, regulate the conformation and distribution of amphiphiles and membrane proteins, including ion-translocating proteins like voltage-gated ion channels, pumps, and carriers embedded in the lipid bilayer matrix. Such field-assisted structural regulation enables the selective transport of materials across the lipid bilayer [14], and advocates the clustering of membrane-anchored proteins [15]. These ordered domains of proteins with larger dipole potentials significantly impact the sorting of membrane-associated molecules and cell signaling properties of the membranes [13,15].

The 2D nanocomposite system examined in this study is composed by highly polarizable nano-inclusions embedded in a fluid and exposed to a perpendicular electric field. Employing molecular dynamics simulations, we investigate the perpendicular (out-of-plane) polarization of the 2D nanocomposite. Under low electric fields, where the induced dipole moments of the inclusions align parallel to the applied field, the inclusions repel each other, leading to random dispersion in the film. Notably, at high electric fields, a qualitatively distinct behavior emerges: the entire system spontaneously organizes into a state that incorporates both parallel and antiparallel polarizations with a fraction of the inclusions counter-polarized to the electric field direction. This unexpected counter-polarization is attributed to the presence of neighboring dipoles that are highly polarized and locally reverse the total electric field's direction. Due to the attractive nature of dipoles with opposite moments, the system exhibits a diverse range of novel equilibrium structures, influenced by inclusion density and electric field strength. These structures encompass fluids and solids with homogeneous polarizations, as well as equilibrium clusters and demixed states with two distinct polarization signatures. Our findings are based on a linearized polarization model that incorporates effective many-body interactions between the inclusions, solved self-consistently using a hybrid technique. These results provide insights for controlling nanocomposites in various applications, such as sensing external fields, directing light within plasmonic materials, and influencing the functionality of biological membranes.

We emphasize here that the demixing in our system differs qualitatively from previous work of phase separation for dipolar systems in three [16–22] or two dimensions [6,17,23–39]. In all of these previous works, the dipolar moments of the particles were either fixed or determined by the external field alone but were never self-consistently determined from neighboring polarizable particles which is the important ingredient of the present work. A similar idea of self-consistent polarization was brought up and modeled by Wilson and Madden [40–43] but this was done in the context of ionic metals in three dimensions and not in two-dimensional layers where the counter polarization effects are more pronounced.

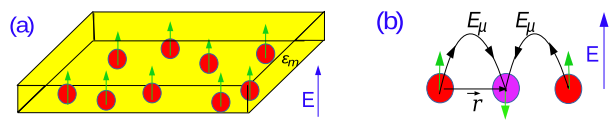

**Fig. 1.** (a) Schematic representation of the system set-up. Polarizable inclusions, shown as red spheres, are suspended in a background dielectric liquid, shown in yellow, and the whole system is subjected to an external field $\vec{E}$ oriented perpendicular to the film surface. Green arrows indicate induced dipole moments of inclusions. (b) A simplified representation of the associated dipole triplet. The central inclusion at the position $\vec{r}$ develops a negative dipole moment when the dipolar field of the neighbors, $2|\vec{E}_\mu(\vec{r})| > E$, resulting in a local counter-polarization. For details see the text and Appendix A. (For interpretation of the colors in the figure(s), the reader is referred to the web version of this article.)

The rest of this paper is organized as follows. Our simulation model for the 2D nanocomposite film with mobile and polarizable inclusions is described in section 2. In section 3 we provide a comprehensive overview of our molecular dynamics simulations. Section 4 encompasses simulation outcomes. Here we showcase simulation snapshots, deliberate on dipole association triplet and dipole clustering effects, and conduct an analysis of dipole moment distribution functions and structure factors. Finally, our conclusions are presented in section 5.

## 2. The model

We consider a 2D nanocomposite film comprising $N$ polarizable high-$k$ inclusions of dielectric permittivity $\varepsilon_p$ and hard core diameter of $\sigma$ suspended in a fluid background of a lower permittivity $\varepsilon_m < \varepsilon_p$ at positions $\vec{r}_i$ ($i=1,...,N$) as depicted in Fig. 1. The film thickness is fixed to $L_z = \sigma$, and its lateral dimensions $L_x = L_y = L$ are determined by the 2D packing fraction of inclusions $\eta = N\pi \left[\sigma/(2L)\right]^2$. Under an applied field $\vec{E}$, perpendicular to the film surface, the particles gain an induced dipole moment $\vec{\mu}_0 = \alpha \vec{E}$ parallel to $\vec{E}$, resulting in repulsive dipole-dipole interactions between them.

We model the situation with a Hamiltonian that depends on the configuration of the inclusions but also parametrically on their actual dipole moments $\{\vec{\mu}_i\}, i=1,...,N$. The dipole moments adjust themselves instantaneously to the momentaneous configuration such that they minimize the Hamiltonian for fixed inclusion coordinates. At the minimum, a situation is realized where the polarization is selfconsistently and uniquely governed by the local electric field produced both externally and by the neighbors. In other terms, the dipole moments $\{\vec{\mu}_i\}, i=1,...,N$ follow adiabatically the dynamical positions of the inclusions. Thermal fluctuations of the polarized dipoles are neglected. We remark that this is formally similar to the Born-Oppenheimer approximation in solid state physics where the electron density follows adiabatically the ion dynamics [44]. In practice this is resulting in effective many-body forces [45–47] induced via the polarization acting on the inclusions.

In detail the Hamiltonian for the considered system is written as,

$$H\left(\{\vec{r}_i\},\{\vec{p}_i\},\{\vec{\mu}_i\},i=1,....,N\right)$$
$$= \sum_{i=1}^{N} \frac{p_i^2}{2M} + \sum_{i=1}^{N} \sum_{\substack{j=1, j\neq i}}^{N} U_{ij} - \sum_{i=1}^{N} \mu_i E + \sum_{i=1}^{N} U_i \qquad (1)$$

where $\{\vec{r}_i\}$, $\{\vec{p}_i\}$ and $\{\vec{\mu}_i\}$ denote the whole coordinate, momentum, dipole moment sets for the inclusions, and $M$ is the inclusion mass. Here the first term is the total kinetic energy of the in-plane movement of inclusions resulting in a finite temperature of the system, and the second term is the total potential energy of the dipolar interactions between all inclusions,

$$U_{ij} = \begin{cases} \infty, & \text{for } r_{ij} \leq \sigma, \\ \frac{\mu_i \mu_j}{4\pi\varepsilon_0 r_{ij}^3}, & \text{for } r_{ij} > \sigma, \end{cases} \qquad (2)$$

where $\varepsilon_0$ denotes the permittivity of vacuum, and $r_{ij} = |\vec{r}_j - \vec{r}_i|$ represents the distance between the $i$-th and $j$-th induced dipoles. The third term in Eq. (1) represents the potential energy of the induced dipole $\vec{\mu}_i$ under the applied field $\vec{E}$. The fourth term in Eq. (1) represents the self-polarization energy of the inclusion [22,48],

$$U_i = \frac{1}{2\alpha} \mu_i^2 \qquad (3)$$

where the expression for the inclusion polarization coefficient $\alpha$ is given in Appendix A.

By minimizing the Hamiltonian over the induced dipole moment $\mu_i$,

$$\frac{dH(\{\vec{r}\},\{\vec{p}\},\{\vec{\mu}\})}{d\mu_i} = 0 \qquad (4)$$

we arrive at the following equation for $\mu_i$,

$$\sum_{j=1}^{N} \frac{\mu_j}{4\pi\varepsilon_0 r_{ij}^3} - E + \frac{\mu_i}{\alpha} = 0 \qquad (5)$$

which can be rewritten as,

$$\vec{\mu}_i(\vec{r}) = \vec{\mu}_0 + \vec{\mu}_i^{(j)}(\vec{r}) \qquad (6)$$

where

$$\vec{\mu}_i^{(j)}(\vec{r}) = \alpha_d \sum_{j=1, j\neq i}^{N} \vec{E}_j(\vec{r}_i) \qquad (7)$$

denotes the contribution from the combined field of all other dipoles. Eq. (5) is the basic self-consistent condition to determine the actual dipole moments.

Expressions for the polarization coefficient $\alpha_d$ and for the cumulative field $\sum_j \vec{E}_j$ can be found in Appendix A. The field $\vec{E}_j$, strictly speaking, is not homogeneous inside the $i$-th particle at intermediate and high $\eta$, and a volume integration of the neighboring fields $\vec{E}_j$ for the calculation of $\mu_i$ should be carried out [26]. Nevertheless, for the sake of simplicity, we assume $\vec{E}_j$ homogeneous inside the target inclusion.

The amplitude of the dipoles $\mu_i$ and the average separation distance $r_{av}$ between neighboring dipoles define the interaction strength $\Gamma_\mu$ between dipoles,

$$\Gamma_\mu = \frac{\mu_0^2}{4\pi\varepsilon_0 \varepsilon_m r_{av}^3 k_B T} \qquad (8)$$

where

$$r_{av} = \frac{2L}{\sqrt{N\pi}} = \frac{\sigma}{\sqrt{\eta}} \qquad (9)$$

We introduce a coupling parameter for the 2D systems as,

$$\Gamma = \frac{\Gamma_\mu}{\eta^{3/2}} = \frac{\alpha^2 E^2}{4\pi\varepsilon_0 \varepsilon_m \sigma^3 k_B T} \qquad (10)$$

which does not depend on the particle density [16]. When $\Gamma$ is of the order of 1, the correlation effects between the induced dipoles can be ignored. This means that $\vec{\mu}_i = \vec{\mu}_0 = \alpha E$, and as a result, the third part in Eq. (1) becomes $N\alpha E^2$. This is typically observed in magnetically polarizable particles, which develop weak induced dipoles because the magnetic susceptibility of such materials rarely exceeds 1.

For large values of $\Gamma$, the correlation effects are significant, and their impact on individual dipoles $\vec{\mu}_i$ through the many-body term in Eq. (7) is not known beforehand. As we will demonstrate in the following sections, strong correlations could result in dipole clustering and fluid-gas demixing in the system under consideration.

## 3. Details of computer simulations

We employed Langevin Dynamics simulations to study the behavior of $N = 8192$ induced dipoles of diameter $\sigma = 2000$ $A$ in a flat 2D dielectric film under an applied field $E$. Molecular dynamics (MD) runs with a Langevin thermostat with a friction coefficient $\gamma = 2$ $\mu s^{-1}$, and a Gaussian white-noise force were performed in a constant $NVT$ ensemble. The equations of motion were numerically integrated with a time step $\Delta t = 1.0\text{e-}4$ $t_0$, where $t_0 = \sqrt{4\pi\varepsilon_0 M\sigma^3/e^2}$, and $e$ represents the elementary charge. Accounting for an inclusion density of 1 g/cm$^3$, the simulation's time step equates to approximately 1 ns.

The motion of inclusions was confined to the $xy$-plane. Standard periodic boundary conditions were imposed by filling space with translational replicas of the fundamental cell in the $x$ and $y$ directions. The long-range electrostatic interactions between the induced dipoles were handled using the standard Lekner summation algorithm [49].

Initially, all inclusions were randomly distributed at positions $\vec{r}_i$, $i = 1, ... N$, with no overlapping between particles, $|\vec{r}_i - \vec{r}_j| \geq \sigma$. The lateral box size $L$ was varied from $358\sigma$ to $146\sigma$ to adjust the packing fraction from $\eta = 0.05$ to $\eta = 0.3$. This upper limit is much below the maximum 2D random packing for circles [50,51] $\eta_r = 0.82\text{-}0.83$, as well as the highest packing of circles on a plane $\eta_m = \pi/(2\sqrt{3}) = 0.907$.

The applied field was varied within the range of $0.01E_0$ to $50E_0$, where $E_0 = 1$ MV/m, allowing us to adjust the coupling parameter $\Gamma$ between $10^{-3}$ and 8000. The simulation parameters for the 9 runs, each corresponding to a specific value of the applied field $E$ and consisting of a series of 7 simulations with different $\eta$, are summarized in Table 1.

The simulation procedure is a hybrid of combining the determination of the dipole moments and the motion of the inclusions [52]. In detail, it consists of a series of iterations at each time step to stabilize the inclusion dipoles $\vec{\mu}_i$ self-consistently. Initially, a dipole moment $\vec{\mu}_0 = \alpha \vec{E}$ is assigned to all inclusions. Subsequently, during the first iteration, at step $k = 1$, the value of $\vec{\mu}_i^{(j)} = \varepsilon_m \alpha \sum_{j \neq i}^{N} \vec{E}_j(\vec{r}_i, \vec{\mu}_0)$ is computed for all inclusions $i$. Here, the field $\vec{E}_j(\vec{r}_i, \vec{\mu}_0)$ is determined by Eq. (25) in Appendix A, and $\mu_0$ in its argument indicates that the neighboring $j$-th inclusion possesses a dipole moment $\vec{\mu}_0$. Subsequently, the total dipole moment

$$\vec{\mu}_i|_{k=1} = \vec{\mu}_0 + \varepsilon_m \alpha \sum_{j \neq i}^{N} \vec{E}_j(\vec{r}_i, \vec{\mu}_0) \qquad (11)$$

is evaluated for all inclusions at the iteration step $k = 1$. In the subsequent iteration step $k = 2$, Eq. (11) is reiterated for all inclusions, as follows,

$$\vec{\mu}_i|_{k=2} = \vec{\mu}_0 + \varepsilon_m \alpha \sum_{j \neq i}^{N} \vec{E}_j(\vec{r}_i, \vec{\mu}_i|_{k=1}) \qquad (12)$$

**Table 1**
Simulation parameters for runs *1–9*. Each run consisted of 7 simulations corresponding to $\eta = 0.05, 0.1, 0.15, 0.2, 0.25, 0.28, 0.3$.

| Runs | 1 | 2 | 3 | 4 | 5 | 6 | 7 | 8 | 9 |
|---|---|---|---|---|---|---|---|---|---|
| $E/E_0$ | 0.01 | 0.1 | 0.2 | 0.5 | 1 | 2 | 5 | 10 | 50 |
| $\Gamma$ | $10^{-3}$ | 0.03 | 0.1 | 1 | 3 | 12 | 76 | 300 | 8000 |

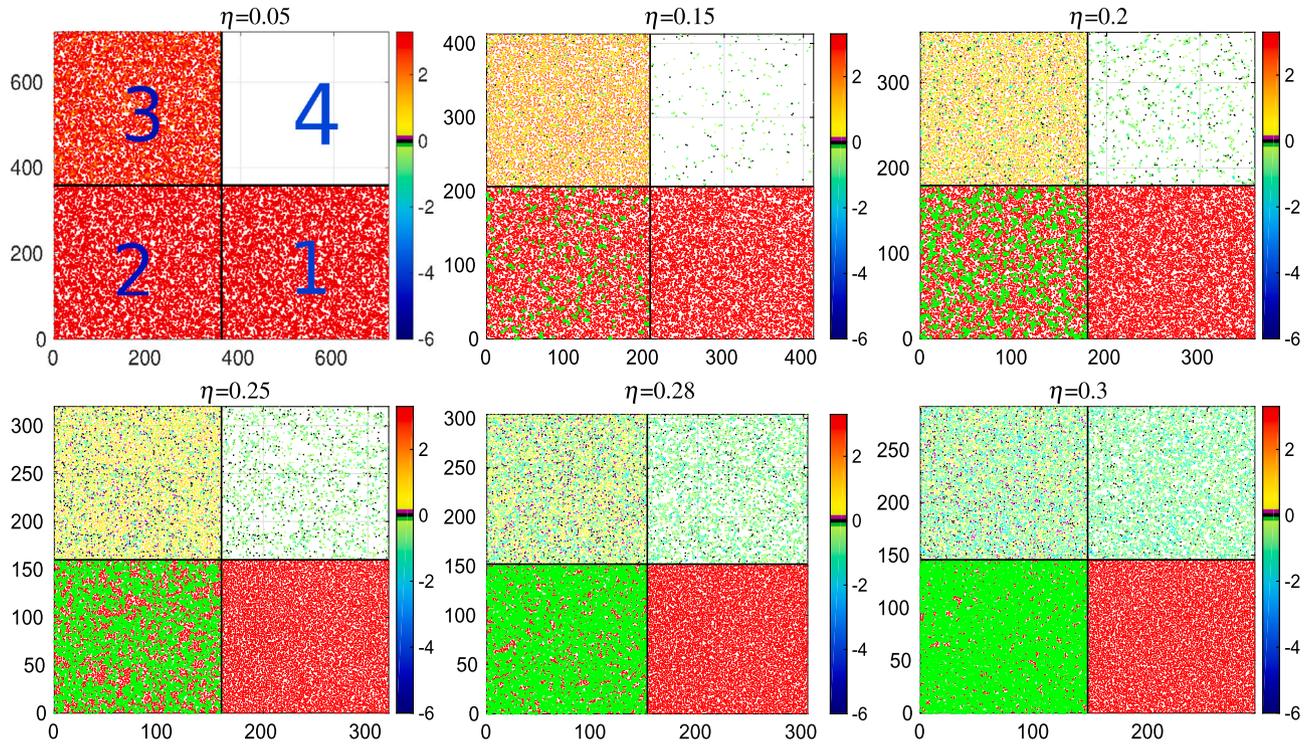

**Fig. 2.** Simulation snapshots for the weak-field run *5* with $E = E_0$ and $\mu_0 = 3.3 e\sigma$. The labels from 1 to 4 in the snapshot for $\eta = 0.05$ indicate four sections with different levels of the dipole association and orientation information. For this case with no counter-polarized dipoles, the sections 1, 2, and 3 look alike.

After performing certain mathematical manipulations, this equation can be expressed as,

$$\vec{\mu}_i|_{k=2} = \vec{\mu}_i|_{k=1} + \varepsilon_m \alpha \sum_{j \neq i}^{N} \vec{E}_j(\vec{r}_i, \vec{\mu}_i|_{k=1}) - \varepsilon_m \alpha \sum_{j \neq i}^{N} \vec{E}_j(\vec{r}_i, \vec{\mu}_i|_{k=0}) \quad (13)$$

where $\vec{\mu}_i|_{k=0} = \vec{\mu}_0$.

The iterations continue until step $k = n$,

$$\vec{\mu}_i|_n = \vec{\mu}_i|_{n-1} + \varepsilon_m \alpha \left( \sum_{j \neq i}^{N} \vec{E}_j(\vec{r}_i, \vec{\mu}_i|_{n-1}) - \sum_{j \neq i}^{N} \vec{E}_j(\vec{r}_i, \vec{\mu}_i|_{n-2}) \right) \quad (14)$$

provided that the difference between the cumulative sums becomes negligible. The stabilized induced dipole for inclusion $i$ is then assumed to be $\vec{m}_i = \vec{\mu}_i|_n$.

For highly dilute systems with small $\Gamma \ll 1$, one can assume that the stabilized dipoles are the same, $\vec{m}_i = \vec{m}$. Then, the iterative procedure leads to a simple expression for the dipole $\vec{m}$, as shown in Appendix B. For the non-dilute systems with moderate values of $\Gamma$, the stabilized dipoles $\vec{m}_i$ are different. Therefore, the considered 2D nanocomposite can be assumed as a collection of independent dipoles $\{\vec{m}_i\}$ at the simulation time $t$. Then, the force $\vec{F}_i(r_i)$ acting on the $i$-th inclusion is calculated as,

$$\vec{F}_i(r_i) = -\frac{dH(\{\vec{r}\}, \{\vec{p}\}, \{\vec{\mu}\})}{dr_i} = -\frac{3}{4\pi\varepsilon_0 \varepsilon_m} \sum_{j \neq i}^{N} \frac{m_i m_j}{r_{ij}^5} \vec{r}_{ij} \quad (15)$$

This is followed by updating the positions of all inclusions, $\vec{r}_i$, using Langevin Dynamics moves. The process then proceeds to a new set of iterations for dipole stabilization at the next simulation time step.

In total, the system was first equilibrated under the applied field during $N_e = 10^6$ simulation steps. Then, another $N_p = 3 \cdot 10^6$ simulation steps were used as production runs to calculate the needed quantities.

### 4. Simulation results

#### 4.1. Weak coupling, local counter-polarization, and dipole association effects

In this section we will analyze snapshots from the run *5* which corresponds to weak coupling $\Gamma = 3$. Snapshots for other weak coupling $\Gamma \leq 1$ runs *1–4* from Table 1 are relocated to the Supplementary Information (SI) because they exhibit similar characteristics as the run *5*.

Each snapshot in Fig. 2 is divided into four sections, highlighting the particle distribution based on their respective dipole information.

– The *bottom right section*, referred to as section 1, portrays a standard snapshot wherein the inclusions are illustrated as dispersed red dots.

– The *bottom left section*, referred to as section 2, features the same red dots, while additionally highlighting configurations of "associated dipole" triplets by enclosing them with green circles. Each associated dipole triplet, as depicted schematically in Fig. 1b, comprises a counter-polarized dipole, i.e. here a negative dipole oriented against the applied field due to the influence of negative dipolar fields from neighboring dipoles (refer to Eq. (25) in Appendix A for specifics). In Appendixes C, D, and E, various configurations are discussed, particularly those with critical $\eta_c$ where the negative fields of neighboring induced dipoles can counteract the applied field $E$. Elongated inclusions, however, when their c-axis is aligned with the applied field, may induce dipole moment reversal of the target inclusion even at low packing fractions.

– In the *top left section*, referred to as section 3, each inclusion is additionally color-coded based on the direction and magnitude of its dipole moment $\vec{m}_i$. The corresponding color-bar is provided on the snapshot's right side for reference. These dipole moments are scaled as $\mu_i^* = m_i/(e\sigma)$, and the scaled induced dipole from the applied field is denoted as $\mu_0^* = \alpha E / (4\pi\varepsilon_0 e\sigma)$, where $E$ is measured in [V/m] units.

– Finally, in the *top right section*, referred to as section 4, only the inclusions with negative dipoles $\vec{m}_i < 0$ are depicted, aiding in the identification of regions demonstrating diamagnetic behavior.

In Fig. 2, at low $\eta = 0.05$, as observed in section 1, all dipoles are randomly dispersed throughout the system. Due to the low value of the coupling $\Gamma$, interactions remain weak, causing the inclusions to mimic behavior akin to a hard sphere fluid. This accounts for the occasional formation of cluster-like structures, albeit these structures are transient and lack stability.

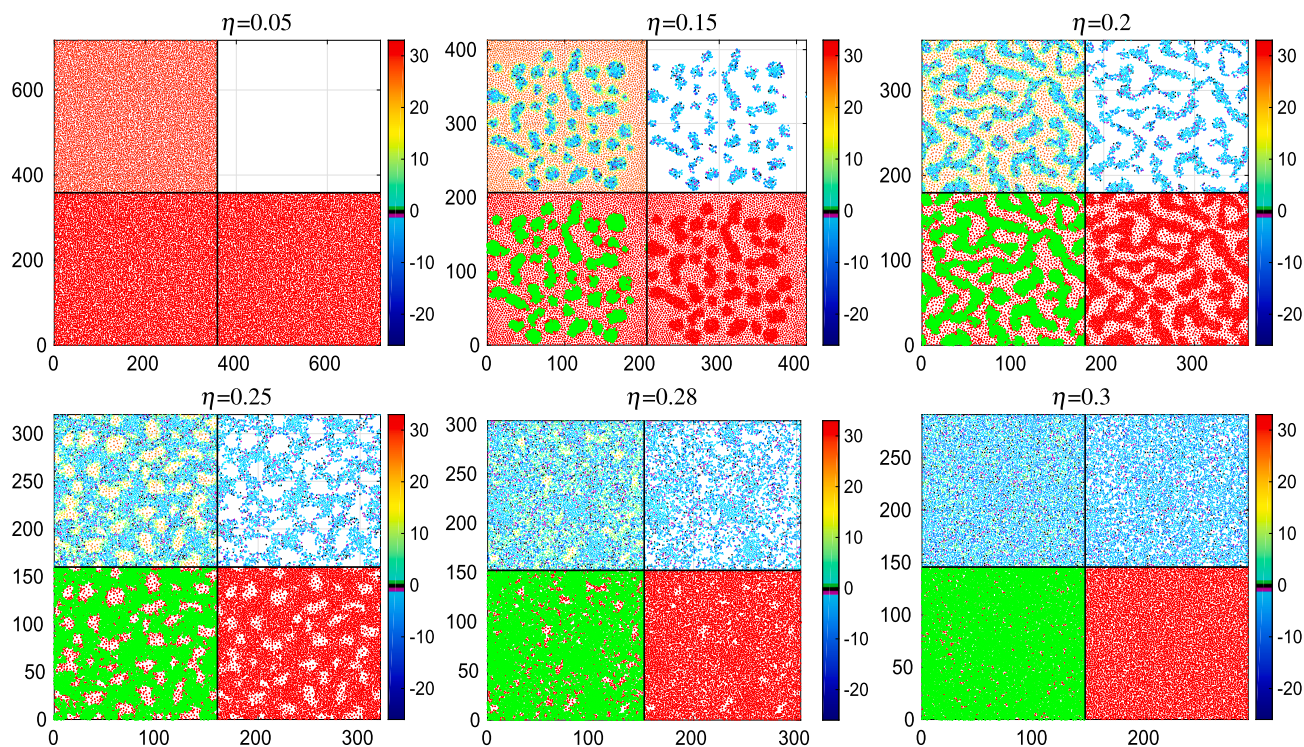

**Fig. 3.** Simulation snapshots for the strong-field run *8* with $E = 10 E_0$ and $\mu_0 = 33 e\sigma$.

Section 2 exhibits no associated dipoles whatsoever. In Section 3, the coloring of inclusions distinctly reveals that all dipoles are positive, albeit their values are smaller than $\mu_0$. This reduction stems from the collective suppressing effect induced by the dipolar field of neighboring inclusions (for specifics, refer to Eq. (25) in Appendix A). Section 4 is empty because all induced dipoles register as positive.

As $\eta$ increases, there is a corresponding rise in the count of associated dipoles, as depicted by the green shading in section 2 for $\eta = 0.1$–0.3. In these packing fractions, the overall coloring in section 3 gradually transitions from red/orange to yellow/blue, indicating a decrease in all dipole values. Simultaneously, the quantity of negative dipoles shown in section 4 also increases alongside the increment in $\eta$. At $\eta = 0.2$, nearly half of the inclusions form associations, while at $\eta = 0.3$ this proportion increases to about 95%, and half of the inclusions acquire negative induced dipoles.

### 4.2. Strong coupling and dipole clustering effects

Simulation snapshots for the strong-field scenario $E = 10 E_0$, run *8* with $\Gamma = 300$, are presented in Fig. 3. Snapshots from other strong-field runs *6*, *7*, and *9* with $\Gamma > 10$ from Table 1 share analogous characteristics to run *8* and have consequently been relocated to the SI.

At low $\eta \leq 0.1$, dipole association doesn't occur due to the robust repulsion between induced dipoles. However, when $\eta \geq 0.15$, associated triplets begin to form clusters, leading to a fluid-gas demixing within the system. In the snapshot at $\eta = 0.15$ in Fig. 3, section 2 illustrates that these clusters possess a compact, rounded shape. Section 3 of the same snapshot reveals a complex distribution of dipoles, showing a continuous transfer of dipole orientation from the strongly positive background to the less positive cluster edges and eventually to the negative cluster centers displaying diamagnetic behavior. The density of particles within the cluster is higher than in the background.

With increasing $\eta$, the clusters amalgamate, creating a continuous stripe-like network primarily composed of negative dipoles, evident in the snapshot for $\eta = 0.2$, section 2. These networks contain pockets filled with positive dipoles. Further elevation of $\eta$ to 0.25 generates bubbles occupied by very low-density positive dipoles amidst a background of densely populated associated dipoles.

The clustering patterns observed in Fig. 3, and in Figs. 6 (run *7*) and 7 (run *9*) in SI, for associated dipoles at $\eta = 0.15$ and 0.2 exhibit similarities to microphase separation phenomena found in systems characterized by competing interactions: specifically, a combination of long-range repulsion and short-range attraction. Such systems include diblock copolymers [53–56], 2D colloidal systems [57,58], biological macromolecules featuring long-range Coulomb repulsion and short-range solvent-induced attraction [59], binary mixtures of oppositely charged particles with additional short-range attraction among like particles and short-range repulsion between different ones [60], mixed polymer brushes with chain-length asymmetry and solvent selectivity [61], and fluid mixtures [62]. In our polarized nanocomposite, these competing interactions arise from the repulsion between positively oriented dipoles and the attraction between oppositely oriented dipoles. Consequently, akin to other systems with competing interactions, we observe a diverse array of equilibrium microphases, including clusters, stripes, and bubbles, at specific values of the applied field and particle density.

### 4.3. Phase diagram of the polarized nanocomposite on the $(E, \eta)$ plane

The comprehensive phase diagram on the $(E, \eta)$ plane, shown in Fig. 4, delineates five distinct states represented by various geometric symbols.

1. A magenta square represents states characterized by low $\eta$ and absence of dipole association, thus featuring entirely positively oriented dipoles randomly distributed within the host matrix.

2. A blue circle represents states at intermediate $\eta$, which house partially associated and counter-polarized dipoles randomly distributed within the host matrix.

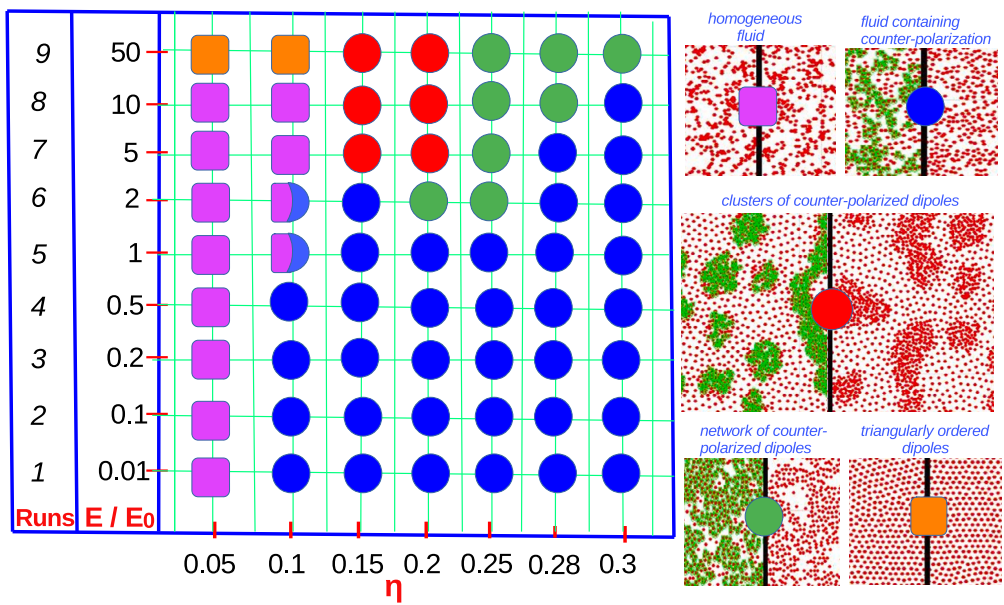

**Fig. 4.** The predicted phase diagram of the polarized 2D nanocomposite on the $(\log(E/E_0), \eta)$ plane. Magenta squares are for the classical dipolar system with no dipole-dipole association. Blue circles are for systems with partially associated and counter-polarized dipoles without any clustering. Red circles are for fluid-gas demixed systems featuring high particle density clusters of counter-polarized dipoles surrounded by low density liquid of positively oriented dipoles. Green circles are for systems with voids filled with a few positively oriented dipoles surrounded by a high density crystalline network of counter-polarized dipoles. Orange squares are for homogeneous systems with triangular ordering for positively oriented dipoles.

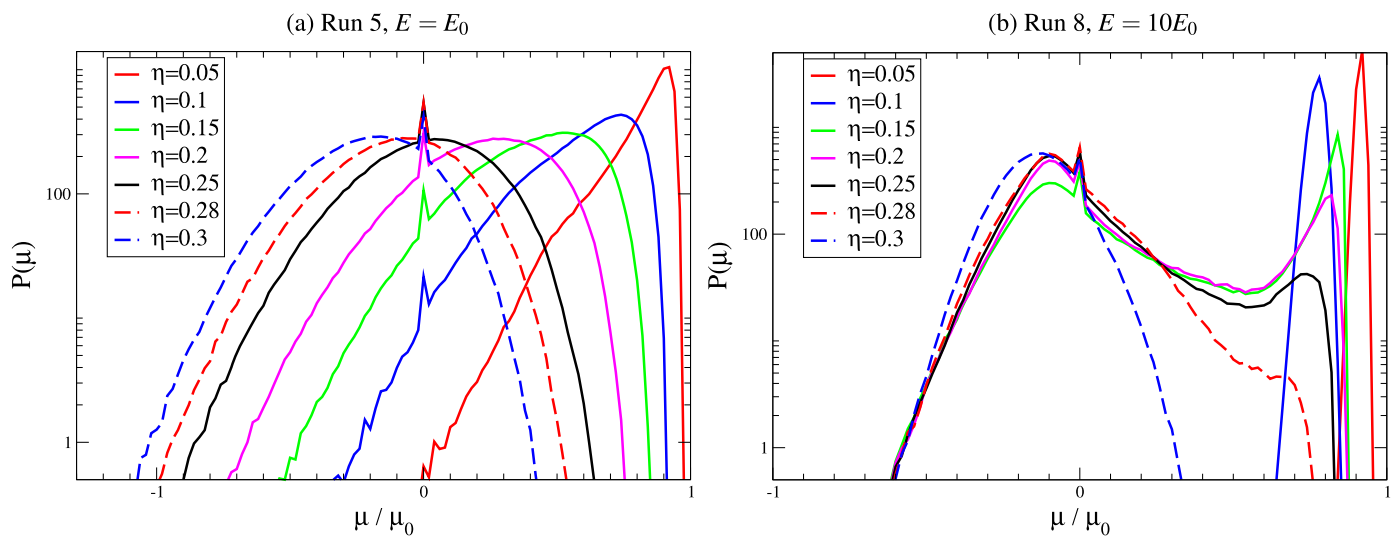

**Fig. 5.** Dipole distribution function $P(\mu)$ for induced dipoles. (a) A weak-field simulation run 5 with $E = E_0$. (b) A strong-field simulation run 8 with $E = 10E_0$. Full lines: red- $\eta = 0.05$, blue- $\eta = 0.1$, green- $\eta = 0.15$, magenta- $\eta = 0.2$, black- $\eta = 0.25$. Dashed lines: red- $\eta = 0.28$, blue- $\eta = 0.3$. Note that the y-axis is on a log scale.

3. A red circle represents highly poled states with $E \geq 5E_0$ and $\eta = 0.15, 0.2$, which feature clusters of counter-polarized dipoles amidst a background of positively oriented dipoles. These states indicate a higher particle density within the clusters compared to the background, a fingerprint of the fluid-gas demixing.

4. A green circle represents other highly poled states with $E \geq 2E_0$ and $\eta \geq 0.2$, where the pockets of the dense network of counter-polarized dipoles are occupied by a few positively oriented dipoles.

5. An orange square represents extremely poled states with $E = 50E_0$ and $\eta \leq 0.1$, which feature a crystalline order for the positively oriented dipoles homogeneously distributed within the host matrix.

### 4.4. Probability distribution of dipole moments

To gain a deeper understanding on the polarization tendencies of inclusions in the nanocomposite, we present in Fig. 5 the probability distribution $P(\mu)$ of induced dipoles for runs 5 and 8. Detailed $P(\mu)$ distributions for additional runs are available in the SI. In the case of a weak-field polarization run 5, depicted in Fig. 5a, several intriguing phenomena unfold. Firstly, with an increase in $\eta$, there's a noticeable shift in $P(\mu)$ towards lower $\mu/\mu_0$ values. This underscores that with a higher coupling parameter $\Gamma$, the influence of neighboring induced dipoles becomes more prominent, resulting in a suppression of the dipole moment within the targeted inclusion. Secondly, the distribution demonstrates asymmetry, particularly at low $\eta$, where the right arm appears sharper in contrast to the left arm. As $\eta$ increases, the symmetry of $P(\mu)$ tends to improve, evident in the transition from the red solid line to the black dashed line. Thirdly, at $\eta = 0.25$, as depicted by the solid black line, the distribution peaks at $\mu \approx 0$, indicating the involvement of approximately half of the particles in associated dipolar triplets.

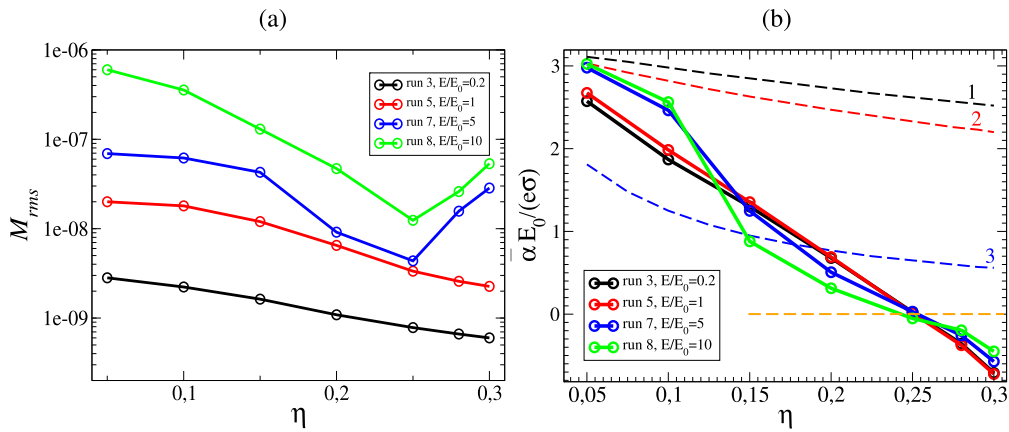

**Fig. 6.** (a) The dependence of the root mean square fluctuation $M_{rms}$ of the total dipole moment **M** of all inclusions, defined in Eq. (16), on the packing fraction $\eta$. (b) The dependence of the renormalized system polarizability $\bar{\alpha}$ per particle, defined in Eq. (19), on the packing fraction $\eta$. Dashed lines represent the analytically predicted $\bar{\alpha}$ as given by Eq. (20): the black line labeled as *1* corresponds to $\beta_p = 1$ and $B = 8$, the red line labeled as *2* corresponds to $\beta_p = 1$ and $B = 13$, and the blue line labeled as *3* corresponds to $\beta_p = 0.1$ and $B = 13$. Further elaboration can be found in the text concluding section 4.5.

In the case of a strong-field polarization run *8*, observed in Fig. 5b, $P(\mu)$ shows a double-peaked structure within the range $0.15 \leq \eta \leq 0.25$. These peaks correspond to the fluid-gas demixing characterized by dense negative clusters, both in compact forms and as a network. In essence, the fluid-gas demixing is evident not only in particle density correlations but also in dipole moment correlations. The right peak in $P(\mu)$ centers around $\mu/\mu_0 \approx 0.85$, while the left peak resides near $\mu/\mu_0 \approx -0.1$. As the packing fraction $\eta$ increases, the right peak diminishes, and the remaining peak shifts towards more negative dipole values.

*4.5. Dipole moment fluctuations and polarizability of the system of induced dipoles*

Fig. 6(a) shows the root-mean-square fluctuation $M_{rms}$ of the mean dipole moment $\bar{M}$ for the low-field runs *3*, *5*, and the strong-field runs *7* and *8*. This function is characterized by:

$$M_{rms} = \left( \frac{1}{\tau} \int_{t_0}^{t_0+\tau} \left[ M(t) - \bar{M} \right]^2 dt \right)^{1/2} \quad (16)$$

where $M(t) = \sum_{i=1}^{N} \mu_i(t)$ is the instantaneous total dipole moment of the system at the simulation time $t$, and $\bar{M} = \int_{t_0}^{t_0+\tau} M(t) dt / \tau$ is the total dipole moment of the system averaged over a time interval of $\tau$ starting from the arbitrary simulation time $t_0$.

The low-field $M_{rms}$ data, represented by black and red lines in Fig. 6 corresponding to runs *3* and *5*, consistently diminish as $\eta$ increases. This pattern aligns with expectations since an elevation in $\Gamma$, following Lindemann's rule, results in decreased positional fluctuations of inclusions, consequently reducing the fluctuation of $\mu_i^{(j)}$ as described in Eq. (6).

Contrarily, the high-field $M_{rms}$ data, represented by the blue and green lines in Fig. 6 corresponding to runs *7* and *8*, exhibit a non-monotonic trend with minima occurring at $\eta = 0.25$. At this particular packing fraction, according to the phase diagram outlined in Fig. 4, the pockets within the high-density cluster network of negative dipoles manifest as low-density voids housing positive dipoles.

The rise in $M_{rms}$ at higher $\eta$ can be elucidated as follows: Let us assume that at some $\eta_a$, the voids occupy a fraction $\theta$ of the system's area, and they completely vanish at a higher $\eta_b > \eta_a$. The average separation between the nearest inclusions at these packing fractions can be mathematically expressed as:

$$r_{av}(\eta_a) \approx \sqrt{\frac{1-\theta}{\eta_a}}, \quad r_{av}(\eta_b) \approx \sqrt{\frac{1}{\eta_b}} \quad (17)$$

If the ratio,

$$R_{av} = \frac{r_{av}(\eta_b)}{r_{av}(\eta_a)} \approx \sqrt{\frac{\eta_a}{(1-\theta)\eta_b}} \quad (18)$$

exceeds one, it is expected that the dipole fluctuations would escalate, resulting in $M_{rms}(\eta_b) > M_{rms}(\eta_a)$. Substituting $\eta_a = 0.25$ and $\eta_b = 0.3$ into Eq. (18), assuming $\gamma = 1/2$ (refer to Fig. 3 snapshot for $\eta = 0.25$), yields $R_{av} \approx 1.1$. This implies that the rise in $r_{av}$ linked to the decrease in the void area at higher $\eta$ is not counterbalanced by its reduction due to the increase in $\eta$.

In Fig. 6(b) we present the polarizability of the system of induced particles calculated as,

$$\bar{\alpha} = \frac{\bar{M}}{NE} \quad (19)$$

The polarizability $\bar{\alpha}$ exhibits a decline with increasing $\eta$, reaching zero at $\eta = 0.25$ for both the low-field runs *3* and *5*, as well as for the high-field runs *7* and *8*. Beyond this packing fraction, more than half of the inclusions form associations. Zero polarizability indicates a balance between dipoles polarized along the applied field and their counter-polarized counterparts. This equilibrium is reflected in the dipole distribution $P(\mu)$ illustrated by the black lines in Fig. 5. For $\eta > 0.25$, the system's polarizability becomes negative ($\alpha < 0$). It's worth noting that at these higher packings, the induced dipolar field from neighboring inclusions becomes markedly non-uniform within the $i$-th particle. This suggests that volume integration of neighboring fields $\vec{E}_j$ is necessary for calculating $\mu_i$, as mentioned in [26]. However, in our current study, we maintain the assumption of homogeneity of $\vec{E}_j$ within the $i$-th inclusion. For a deeper comprehension of negative polarization, more intricate simulations involving volume integration of $\vec{E}_j$ are required.

Under the assumption that all dipoles are weakly interacting with each-other, and thus possess the same stabilized dipole $\bar{\mu}$, using Eq. (26) in Appendix B, Eq. (31) in Appendix C, and Eq. (37) in Appendix D, we arrive at the following analytical expression for the system polarization,

$$\frac{\bar{\alpha} E_0}{e\sigma} = \frac{\mu_0}{1 + \frac{\eta B}{\beta_p}} \quad (20)$$

Here the coefficient $B = 8$ for a 2D system of randomly distributed inclusions (see Appendix C), and $B = 13$ for the triangular lattice of particles (see Appendix D). The inclusion shape parameter $\beta_p = 1$ for spherical particles and $\beta_p < 1$ for the elongated particles along the applied field direction. In Fig. 6(b) the dashed black line *1* corresponds to the analytically predicted $\bar{\alpha}$ for $B = 8$ and $\beta_p = 1$, the dashed red line *2* corresponds to $\bar{\alpha}$ for $B = 13$ and $\beta_p = 1$, and the dashed blue line *3* corresponds to $\bar{\alpha}$ for $B = 13$ and $\beta_p = 0.1$. In comparison to analytically predicted polar-izability, the inclusion of dipole-dipole correlations through iterations (Eq. (11)–(14)) results in a sharper decline of simulated $\bar{\alpha}$. It's worth mentioning that within the range $0.15 \leq \eta \leq 0.2$ where clustering of inclusions occurs, the simulated polarizability $\bar{\alpha}$ remains positive.

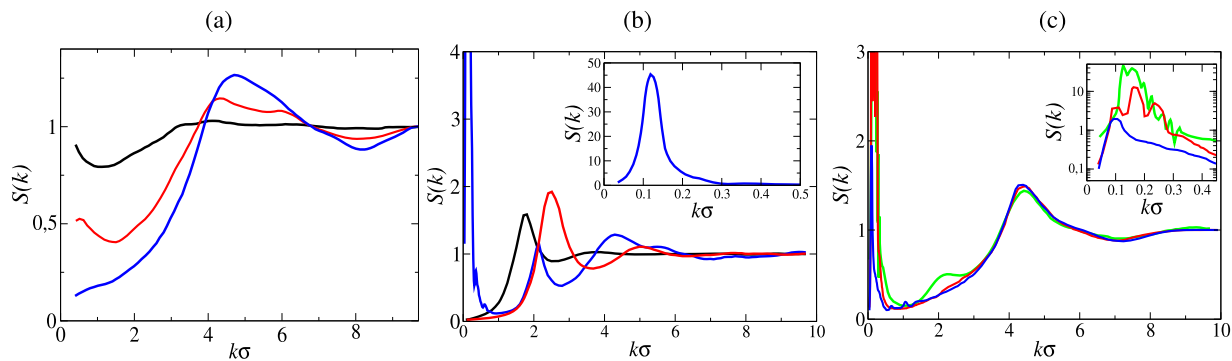

**Fig. 7.** 2D $S(k)$ at different packing fractions $\eta$. (a) $S(k)$ for the low-field run *5* with $E = E_0$, black line- $\eta = 0.05$, red line- $\eta = 0.2$, blue line- $\eta = 0.3$; (b-c) $S(k)$ for the strong-field run *8* with $E = 10E_0$, (b) black line- $\eta = 0.05$, red line- $\eta = 0.1$, blue line- $\eta = 0.15$; (c) green line- $\eta = 0.2$, red line- $\eta = 0.25$, blue line- $\eta = 0.28$.

### 4.6. 2D structure factor and diffraction pictures

The calculated 2D structure factors $S(k)$ for the weak field run 5 are displayed in Fig. 7a. As expected, as $\eta$ increases, the height of the maximum in $S(k)$ rises, and its position shifts to the left. This peak position aligns with the average separation distance $r_{av}$ between neighboring particles, which decreases while $\eta$ increases.

Moving on to the structure factors for the strong field run 8, shown in Fig. 7(b) and (c), a distinct pattern emerges. At $\eta = 0.05$ and $0.1$ (depicted by the black and red lines in Fig. 7(b)), $S(k)$ resembles the structure of strongly correlated liquids. For $\eta = 0.15$ and $0.2$ (blue line in Fig. 7b and green line in Fig. 7c), $S(k)$ exhibits a significant peak at small $k\sigma \approx 0.1–0.3$, consistent with the observed clustering of associated dipoles at these parameters in the snapshots in Fig. 3. The additional maxima at $k\sigma \approx 2$ and $k\sigma \approx 4$ for these packing fractions correspond to the average separations between neighboring particles in the low-density and high-density areas of the cluster network, respectively.

For $\eta = 0.25$ and $0.28$, where the phase diagram predicts the appearance of pockets with low density and positive dipoles, a peak at low-$k$ values persists in $S(k)$, yet the second peak at $k\sigma \approx 2$, which corresponded to the dipole-dipole separation in low density pockets, vanishes. The resemblance between the double-peak appearance in the dipole moment distribution $P(\mu)$ (Fig. 5) and the formation of the low-$k$ peak in $S(k)$ (Fig. 7(b) and (c)) indicates their correlation.

## 5. Discussion

This study unveils several intriguing phenomena observed within 2D nanocomposites containing polarizable inclusions while the dipole-dipole coupling parameter $\Gamma$ was altered in a range from 0.001 to 8000 by varying both the field strength $E$ and the packing fraction $\eta$.

Initially, we observed the formation of associated dipole triplets when the packing fraction reached $\eta \geq 0.1$, across all values of $\Gamma$. These triplets displayed a unique configuration where the dipole moment of the central inclusion opposed the applied field, influenced by the neighboring inclusions' dipolar fields. Each counter-polarized dipole strongly attracted two nearby positive dipoles, rendering these triplets electrostatically stable. Notably, at weak applied fields ($\Gamma \leq 3$), these triplets remained distinct without merging with others. The calculated dipole distribution function $P(\mu)$ in these states displayed a single maximum, shifting towards negative dipole moment values as $\eta$ increased.

Subsequently, under strong applied fields ($\Gamma > 10$) and within a narrow packing fraction range ($0.15 \leq \eta \leq 0.28$), these triplets aggregated into clusters. At $\eta = 0.15$ and $0.2$, compact clusters emerged, hosting high-density counter-polarized dipoles surrounded by a sparse background of positive dipoles. These clusters' boundaries primarily comprised inclusions with minimal to no dipole moments. As the packing fraction rose to 0.28 and 0.3, the compact clusters expanded, ultimately forming a continuous network of counter-polarized dipole clusters. Within this network of high-density inclusions, the pockets displayed a sparse collection of positive dipoles. Notably, the distribution function $P(\mu)$ for these cluster and network states revealed a double-peaked structure. One peak represented the background or pocket dipoles at positive $\mu$ values, while the other peak corresponded to the clustered inverted dipoles or the cluster network at negative $\mu$ values.

In cases where the inclusion retains a constant polarization aligned with the applied field, represented by the permanent dipole vector $\vec{\mu}^p$, the Hamiltonian's second and third terms in Eq. (1) will integrate $\vec{\mu}_i + \vec{\mu}^p$ rather than $\vec{\mu}_i$. Hence, the dipoles $\vec{\mu}_i + \vec{\mu}^p$ will undergo stabilization throughout the iterations outlined in Eqs. (11) to (14). To counter-polarize these dipoles, higher applied fields are necessary. Consequently, as the dipole $\mu^p$ increases, the counter-polarization effect weakens, leading to decreased clustering of inclusions.

In instances where 2D surfaces are not ideal, such as those with undulations, the resulting structures depicted in Fig. 4 might be altered. The regions surrounding peaks and troughs on such surfaces will cause fluctuations in the elevation of the inclusions, potentially fostering attraction between induced dipoles. Consequently, areas atop hills and in the depths of valleys may exhibit lower inclusion density compared to the connecting regions between them. If the undulations are minimal, the phase diagram depicted in Fig. 4 may persist. However, under other circumstances, more intricate structures are anticipated to emerge.

It's noteworthy that, beyond its applications in sensing and energy harvesting, and in understanding the protein conformations in cell membranes, as previously discussed in the introduction, the control of inclusion clustering holds promise for manipulating plasmonic couplings, especially in the context of long-distance propagation of surface plasmon polaritons (SPP) [63–65]. For instance, closely arranged linear chains of silver nanoparticles have been shown to facilitate extended SPP propagation over significant distances, benefiting from minimal radiative losses. In contrast, continuous silver nanowires encounter limitations in SPP propagation distance due to higher losses [66,67].

We expect that the 2D nanocomposite setup can be experimentally accessible and are confident that it can serve to deepen our understanding of the mechanisms behind microphase separation due to competing interactions. The primary concern in the experiments would be the elimination of any non-dipolar interactions, such as the attractive van der Waals forces, among the inclusions. This can be achieved by enveloping them with inert shells [68].

As a future extension of the current study, we plan to use core-shell particle models for higher packing fraction nanocomposites. The iterative dipole stabilization approach used in the current study becomes slow at $\eta \geq 0.33$ and suffers from strong dipole fluctuations. Therefore, to decrease these fluctuations, which mostly stem from the huge dipole moments of the tightly clustered inclusions, each inclusion (a core particle) will be wrapped by an inert shell with $\varepsilon_{shell} = \varepsilon_m$. These shells will not allow the particles to strongly attach to each other in associated dipolar structures, and thus will decrease the fluctuations at higher $\eta$ values.

The inclusion of hydrophilic and hydrophobic interactions between the inclusions, as well as the accounting of their permanent dipole moments is also planned. This level of modeling is necessary for properly analyzing the clustering of dipolar proteins in biomembranes. We also assume explicitly accounting for the inclusion polarization through the poling of its free surface charges. The simple surface charge model, when each dipolar colloid is modeled as two oppositely "charged" particles embedded in a hard sphere [69], or at the edges of the rod [70], might not be sufficient for this purpose. An explicit surface charge model is necessary for higher inclusion concentrations, which properly reacts to the in-plane and out of the plane components of the dipolar fields. Such model can also properly describe the screening of dipolar fields in concentrated systems. Recent studies of screening in concentrated ionic fluids and electrolytes revealed underscreening of electrostatic correlations [71–74]. Despite the enormous effort in performing large-scale simulations and new theoretical investigations, the origin of the anomalously long-range screening length in colloidal systems still remains elusive. Thus, it is believed that more rigorous polarization model is needed at high $\eta$ for poled 2D nanocomposites.

Note that a magnetic field would lead to similar effect within the same model by replacing polarizability with magnetization.

**Supporting information**

Supporting Information is available from the Wiley Online Library or from the author.

**CRediT authorship contribution statement**

**Elshad Allahyarov:** Writing – review & editing, Writing – original draft, Visualization, Validation, Software, Methodology, Investigation, Data curation, Conceptualization. **Hartmut Löwen:** Writing – review & editing, Writing – original draft, Visualization, Validation, Supervision, Resources, Project administration, Methodology, Investigation, Funding acquisition, Conceptualization.

**Declaration of competing interest**

The authors declare that they have no known competing financial interests or personal relationships that could have appeared to influence the work reported in this paper.

**Data availability**

Data will be made available on request.

**Acknowledgements**

The work of H.L. was supported by the DFG within project LO418/23-1. E.A. acknowledges financial support from the Ministry of Science and Higher Education of the Russian Federation (State Assignment No. 075-00270-24-00).

**Appendix A. Electrostatics of nanocomposite polarization**

Let us consider a nanocomposite with a host matrix of an arbitrary form containing polarizable inclusions of arbitrary form placed under external field $\vec{E}$. Each inclusion $i$ then will develop an induced dipole moment $\vec{\mu}_i$ defined as, [22,52,75],

$$\vec{\mu}_i = \vec{\mu}_0 + \vec{\mu}_i^{(j)} = \alpha \vec{E} + \alpha_d \sum_{j \neq i}^{N} \vec{E}_j \tag{21}$$

where the polarization coefficients are given as,

$$\alpha = 3f(\varepsilon_p, \beta_p)\varepsilon_0\varepsilon_m \frac{V_p}{1+(\varepsilon_m-1)\alpha_m^z} \ , \qquad \alpha_d = 3f(\varepsilon_p, \beta_p)\varepsilon_0\varepsilon_m V_p \tag{22}$$

and the permittivity contrast function $f$ is,

$$f(\varepsilon_p, \beta_p) = \frac{\varepsilon_p - \varepsilon_m}{3\varepsilon_m + (\varepsilon_p - \varepsilon_m)\beta_p} \ , \qquad \beta_p = 3\alpha_p^z \tag{23}$$

Here $V_p$ is the inclusion's volume, $\alpha_p^z$ and $\alpha_m^z$ are the depolarization factors for the inclusion and the host matrix along the applied field direction, respectively. Note that $\alpha_d = \varepsilon_m \alpha$ for a flat and infinite host matrix with $\alpha_m^z = 1$ out of the plane direction $z$. The cumulative field $\sum_j \vec{E}_j$ in Eq. (21) accounts for the total dipolar field of the surrounding induced dipoles $\vec{\mu}_j$ at the position of the $i$-th dipole,

$$\vec{E}_j(\vec{r}_i) = \frac{1}{4\pi\varepsilon_0\varepsilon_m} \frac{1}{r_{ij}^3} \left( \frac{3}{r_{ij}^2} \left( (\vec{\mu}_j \vec{r}_{ij}) \vec{r}_{ij} - \vec{\mu}_j \right) \right) \tag{24}$$

where $\vec{r}_{ij} = \vec{r}_j - \vec{r}_i$ is the distance between the two inclusions. For the 2D nanocomposite placed perpendicular to the applied field, Eq. (24) reduces to,

$$\vec{E}_j(\vec{r}_i) = -\frac{\mu_j}{4\pi\varepsilon_0\varepsilon_m r_{ij}^3} \tag{25}$$

The negative sign of this field means that the dipolar fields of neighbors suppress the dipole moment of the target particle.

**Appendix B. Simplified dipole stabilization procedure for very dilute systems**

Under the assumption that all dipoles are well-separated and weakly interact with each-other, and thus possess the same stabilized dipole moment $\vec{m}_i = \vec{m}$, Eq. (11) can be reformulated as,

$$\vec{m} = \vec{\mu}_0 - \vec{m}\frac{\alpha}{4\pi\varepsilon_0}\sum_{j \neq i}^{N} \frac{1}{r_{ij}^3} \ , \quad \vec{m} = \frac{\vec{\mu}_0}{1 + \frac{\alpha}{4\pi\varepsilon_0}\sum_{j \neq i}^{N} \frac{1}{r_{ij}^3}} \tag{26}$$

However, this low-$\eta$ approach proves ineffective in densely packed systems. Take, for instance, a line of inclusions with positions $\vec{r}_i = i\sigma\vec{l}$, where $i = 0, \pm 1, \pm 2, \dots \pm n$, and $\vec{l}$ represents the unit vector along the line. In this scenario, Eq. (26) is transformed as described in [22],

$$\vec{m} = \frac{\vec{\mu}_0}{1 + \gamma \frac{C_A}{\beta_p}} \tag{27}$$

where $C_A \approx 1.2$ is the Apery's constant, and $\beta_p$ denotes for the inclusion shape factor, see Appendix A. For spherical particles $\beta_p = 1$, whereas for elongated particles $\beta_p < 1$. Additionally, the orientation parameter takes values of $\gamma = 1/4$ for $\vec{E}$ perpendicular to $\vec{l}$, and $\gamma = -1/2$ for $\vec{E}$ parallel to $\vec{l}$. Consequently, for slightly elongated inclusions with $\beta_p = C_A/2$, this equation results in induced dipoles that tend towards infinity. To avoid this situation, the iterative dipole stabilization method, outlined in Eqs. (11)–(14), was implemented for the runs 1–9 from Table 1.

**Appendix C. Electrostatic field at the position of the $i$-th inclusion**

Total electrostatic field at position of the $i$-th inclusion, with the help of Eq. (25) can be written as,

$$\vec{E}_i(\vec{r}_i) = \vec{E} + \varepsilon_m \sum_j^{j \neq i} \vec{E}_j(\vec{r}_i) = \vec{E} - \sum_{j=1}^{j \neq i} \frac{\vec{\mu}_j}{4\pi\varepsilon_0 r_{ij}^3} = \vec{E}\left(1 - \frac{\alpha}{4\pi\varepsilon_0}\sum_{j=1}^{j \neq i} \frac{1}{r_{ij}^3}\right) \tag{28}$$

The sum in Eq. (28) can be replaced by the integration as,

$$\sum_{j=1}^{N_j} \frac{1}{r_{ij}^3} = \int_\sigma^\infty \frac{1}{r^3} dN = \int_\sigma^\infty \frac{1}{r^3} \frac{8\eta r dr}{\sigma^2} = \frac{8\eta}{\sigma^3} \quad (29)$$

where we took into account that the number of dipoles $dN$ in the circular strip of area $dS = 2\pi r dr$ around the target dipole $\mu_i$ can be written as,

$$dN = ndS = n2\pi r dr = \frac{8\eta r dr}{\sigma^2} \quad (30)$$

where $n$ is the density of the inclusions $n = N/S = 4\eta/(\pi\sigma^2)$. In Eq. (29) the hard core of the inclusions is taken into account only around the target inclusion. For the second term in Eq. (28) we have,

$$\frac{\alpha}{4\pi\varepsilon_0} \sum_{j=1}^{N_j} \frac{1}{r_{ij}^3} = \frac{\alpha}{4\pi\varepsilon_0} \frac{8\eta}{\sigma^3} = \frac{2\alpha\eta}{\pi\varepsilon_0\sigma^3} \quad (31)$$

If this term is greater than one, then the total field $\vec{E}_i(\vec{r}_i)$ becomes negative, and, as a consequence, the induced dipole $\vec{\mu}_i$ will *counter polarize* and become anti-parallel to the applied field $\vec{E}$.

For a flat nanocomposite with $\alpha = 3\varepsilon_0 f V_p$ and high dielectric contrast $\varepsilon_p \gg \varepsilon_m$ (which is equivalent to $f = 1/\beta_p$), assuming $V_p \approx 4\pi\sigma^3/24$, the *counter-polarization* condition can be written as,

$$\beta_p < \eta \quad (32)$$

This is an interesting result: for spherical particles with $\beta_p = 1$, no counter-polarization is expected at any packing fraction $\eta \leq \eta_m$ for 2D triangular lattice of particles. However, for the particles with $\beta_p < 1$, elongated along the applied field, such counter-polarization is possible. Also, the lower the packing fraction $\eta$, the more elongated should the inclusions be for such counter-polarization to happen.

## Appendix D. A correction to the electrostatic field at the position of $i$-th dipole

In Appendix C, in Eq. (29) the hard core of the inclusions was taken into account only around the target inclusion. At high packing fractions $\eta$ this approach can be corrected by accounting the coordination shells with finite number of neighbors in them [76]. It is expected that when the dipole-dipole interaction parameter $\Gamma > 1$, there is a positional ordering of neighbors around the target dipole, i.e. the system enters into an ordered hexagonal phase. As a result, the coordination number $\bar{N}_k$ of the $k$-th shell ($k = 1, 2, ..., m$) around the target inclusion,

$$\bar{N}_k = \int_{r_{k-1}}^{r_k} dN = \frac{8\eta}{\sigma^2} \int_{r_{k-1}}^{r_k} r dr = \frac{4\eta}{\sigma^2}(r_k^2 - r_{k-1}^2) \quad (33)$$

where $r_k$ is the radius of the $k$-th shell, and $r_0 = 0$, will diverge from the number of particles $N_k$ discussed in Appendix C, especially in the low $k$ shells. For example, for the hexagonal ordering we get $\bar{N}_1 = 6$ against $N_1 \approx 3.64$, $\bar{N}_2 = 6$ against $N_2 \approx 7.28$, and $\bar{N}_3 = 6$ against $N_3 \approx 3.64$. Therefore, for the high dipole-dipole interactions, the summation in Eq. (31) should be corrected by taking, at least, several coordination shells in the hexagonal-like ordering. We here consider the first 8 coordination shells as having hexagonal ordering with the lattice constant $a\sigma$, and treat the rest of the summation in Eq. (31) within the homogeneous distribution approach,

$$\frac{\alpha}{4\pi\varepsilon_0} \sum_{j=1}^{N_j} \frac{1}{r_{ij}^3} = \frac{\alpha}{4\pi\varepsilon_0}\left[\sum_{\delta=1}^{8} \frac{k_\delta}{r_\delta^3} + \int_{4a\sigma}^\infty \frac{8\eta}{\sigma^2}\frac{dr}{r^2}\right] = \frac{\alpha}{4\pi\varepsilon_0}\left[\sum_{\delta=1}^{8} \frac{k_\delta}{r_\delta^3} + \frac{2\eta}{a\sigma^3}\right] \quad (34)$$

Putting $r_1 = a\sigma$, $r_2 = \sqrt{3}a\sigma$, $r_3 = 2a\sigma$, $r_4 = \sqrt{7}a\sigma$, $r_5 = 3a\sigma$, $r_6 = \sqrt{12}a\sigma$, $r_7 = \sqrt{13}a\sigma$, $r_8 = 4a\sigma$, and $k_1 = 6$, $k_2 = 6$, $k_3 = 6$, $k_4 = 12$, $k_5 = 6$, $k_6 = 6$, $k_7 = 12$, $k_8 = 6$ into the sum in Eq. (34), it follows,

$$\frac{\alpha}{4\pi\varepsilon_0} \sum_{j=1}^{N_j} \frac{1}{r_{ij}^3} = \frac{\alpha}{4\pi\varepsilon_0}\left[\frac{9.27}{a^3\sigma^3} + \frac{2\eta}{a\sigma^3}\right] \quad (35)$$

For the triangular lattice the lattice side $a$ and the packing fraction $\eta$ are connected as,

$$\eta = \frac{\pi}{(2\sqrt{3}a^2)} \quad (36)$$

Thus, eliminating $a$ in Eq. (35), and again assuming high contrast composite with flat matrix,

$$\alpha = 4\pi\varepsilon_0\sigma^3 \frac{1}{8\beta_p} \quad (37)$$

we have the counter-polarization condition written as,

$$\beta_p < 1.68\eta^{3/2} \quad (38)$$

This equation clearly indicates that, if the dipole-dipole coupling $\Gamma$ is not small, a counter-polarization of the induced dipole is possible for spherical dipoles with $\beta_p = 1$ at the packing fractions $\eta$ larger than the critical packing fraction $\eta_c = 0.71$. For elongated inclusions with $\beta_p < 1$, the critical $\eta_c$ decreases. For example, if $\beta_p = 0.27$, which corresponds to the ellipsoid-shaped inclusions obeying the geometry [77],

$$x^2 + y^2 + \frac{z^2}{3.5^2} = 1 \quad (39)$$

we get $\eta_c = 0.3$ from Eq. (38).

## Appendix E. Electrostatic field at the position of dipole $\vec{\mu}_i$ in triangular lattice

For a perfect 2D triangular lattice with cell size $a\sigma$, the summation in Eq. (31) can be taken by Madelung-type sums [78],

$$\sum \frac{a^3\sigma^3}{r_{ij}^3} = \sum \frac{1}{\left(\sqrt{n^2 + m^2 + 2nm\cos\left(\frac{\pi}{3}\right)}\right)^3}$$
$$= \sum \frac{1}{\left(\sqrt{n^2 + m^2 + nm}\right)^3} = 6\zeta\left(\frac{3}{2}\right)g\left(\frac{3}{2}\right) \quad (40)$$

where

$$\zeta(x) = \sum_{n=0}^\infty \frac{1}{(n+1)^x} \quad (41)$$

and

$$g(x) = \sum_{n=0}^\infty \frac{1}{(3n+1)^x} - \sum_{n=0}^\infty \frac{1}{(3n+2)^x} \quad (42)$$

By carrying the summations in Eqs. (41) and (42) for $x = 3/2$, we get,

$$\zeta\left(\frac{3}{2}\right) = 2.71, \quad g\left(\frac{3}{2}\right) = 0.74 \quad (43)$$

Thus, for the Eq. (31), with the help of Eq. (36), we have,

$$\frac{\alpha}{4\pi\varepsilon_0} \sum_{j=1}^{N_j} \frac{1}{r_{ij}^3} \approx \frac{\alpha}{4\pi\varepsilon_0} \frac{12.03}{a^3\sigma^3} \approx \frac{\alpha}{4\pi\varepsilon_0} \frac{13\eta}{a\sigma^3} \quad (44)$$

Assuming again a flat matrix with high dielectric contrast, we now arrive at the counter-polarization condition for the triangular lattice as,

$$\beta_p < 1.73\eta^{3/2} \quad (45)$$

Compared to Eq. (38), it is clear that for the crystalline structure the counter-polarization condition can happen at lower $\eta$ for spherical particles, $\eta \geq \eta_c = 0.69$.

**Table of Contents**

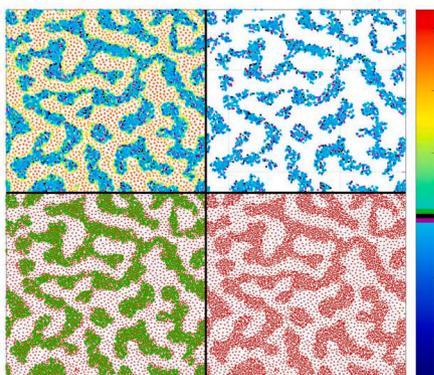

Clustering of induced dipoles in 2D nanocomposite films